\documentclass[%
 reprint,
superscriptaddress,
 amsmath,amssymb,
 aps,
floatfix,
]{revtex4-2}

\usepackage{graphicx}
\usepackage{epsfig} 
\usepackage{dcolumn}
\usepackage{bm}
\usepackage{epstopdf}
\usepackage{multirow}
\usepackage{amsmath}
\usepackage{booktabs}
\usepackage{color}
\usepackage{gensymb}
\usepackage{kantlipsum}  
\usepackage{hyperref}
\usepackage{ulem}
\usepackage{braket}
\usepackage{physics}
\usepackage[utf8]{inputenc}
\usepackage[T1]{fontenc}
\usepackage{mathptmx}

\begin{document}

\title{A High Phase-Space Density Gas of NaCs Feshbach Molecules}

\preprint{APS/123-QED}

\author{Aden Z. Lam}
\affiliation{Department of Physics, Columbia University, New York, New York 10027, USA}
\author{Niccol\`{o} Bigagli}
\affiliation{Department of Physics, Columbia University, New York, New York 10027, USA}
\author{Claire Warner}
\affiliation{Department of Physics, Columbia University, New York, New York 10027, USA}
\author{Weijun Yuan}
\affiliation{Department of Physics, Columbia University, New York, New York 10027, USA}
\author{Siwei Zhang}
\affiliation{Department of Physics, Columbia University, New York, New York 10027, USA}
\author{Eberhard Tiemann}
\affiliation{Institut f\"{u}r Quantenoptik, Leibniz Universit\"{a}t Hannover, 30167 Hannover, Germany}
\author{Ian Stevenson}
\affiliation{Department of Physics, Columbia University, New York, New York 10027, USA}
\author{Sebastian Will}\email{sebastian.will@columbia.edu}
\affiliation{Department of Physics, Columbia University, New York, New York 10027, USA}

\date{\today}

\begin{abstract}
We report on the creation of ultracold gases of bosonic Feshbach molecules of NaCs. The molecules are associated from overlapping gases of Na and Cs using a Feshbach resonance at 864.12(5)$\,$G. We characterize the Feshbach resonance using bound state spectroscopy, in conjunction with a coupled-channel calculation. By varying the temperature and atom numbers of the initial atomic mixtures, we demonstrate the association of NaCs gases over a wide dynamic range of molecule numbers and temperatures, reaching 70~nK for our coldest systems and a phase-space density near 0.1. This is an important stepping-stone for the creation of degenerate gases of strongly dipolar NaCs molecules in their absolute ground state. 
\end{abstract}

\maketitle


The ability to associate weakly bound molecules from ultracold atoms via Feshbach resonances has enabled groundbreaking advances in ultracold quantum science, including the creation of Bose-Einstein condensates (BECs) of Feshbach molecules~\cite{greiner2003emergence, zwierlein2003observation, jochim2003bose, zhang2020atomic}, studies of the BEC-BCS crossover in ultracold Fermi gases~\cite{regal2004observation, zwierlein2004condensation, partidge2005molecular, zwierlein2005vortices}, and the realization of polaronic systems~\cite{schirotzer2009observation,palzer2009quantum,nascimbene2009collective}. Feshbach molecules have also enabled the preparation of ultracold molecules in their absolute ground state~\cite{ni2008high, danzl2010ultracold}. Heteronuclear molecules, which have an electric dipole moment in their ground state, are of particular interest. The long-range interactions between dipolar molecules open up new opportunities for the creation of strongly correlated and highly entangled many-body quantum states~\cite{carr2009cold,lahaye2009physics, baranov2012condensed}, with exciting possibilities for quantum simulation~\cite{micheli2006toolbox,buchler2007strongly,Capogrosso2010} and quantum computing~\cite{demille2002quantum}. 

Many applications of dipolar molecules will require quantum degenerate molecular gases. The coldest molecular gases have been created by associating Feshbach molecules from ultracold atomic mixtures that are then transferred to the rovibrational ground state via stimulated Raman adiabatic passage (STIRAP). So far, ultracold gases of heteronuclear Feshbach molecules have been created with KRb~\cite{ospelkaus2006ultracold,weber2008association}, RbCs~\cite{koppinger2014production,takekoshi2014ultracold}, NaK~\cite{wu2012ultracold}, LiK~\cite{voigt2009ultracold}, NaRb~\cite{wang2015formation}, and LiNa~\cite{heo2012formation}. Typically, numbers have ranged from 1,000 to 20,000 molecules and temperatures from 200 to 1200$\,$nK. Recently, it has been shown that degenerate Fermi gases of KRb~\cite{de2019degenerate} and NaK~\cite{duda2021transition} can be directly created through optimized Feshbach association from the initial Bose-Fermi mixture, reaching temperatures of $50\,$nK and $100\,$nK, respectively. The equivalent for heteronuclear bosonic molecules, the creation of a molecular BEC, has not been achieved yet. For Bose-Bose mixtures enhanced loss in the vicinity of interspecies Fesh\-bach resonances tends to induce heating~\cite{yurovsky1999atom,petrov2004weakly} which so far has limited phase-space densities (PSDs) for bosonic molecular gases to around 0.01~\cite{koppinger2014production,he2021observation}.

\begin{figure} [t]
    \centering
    \includegraphics[width = 8.6 cm]{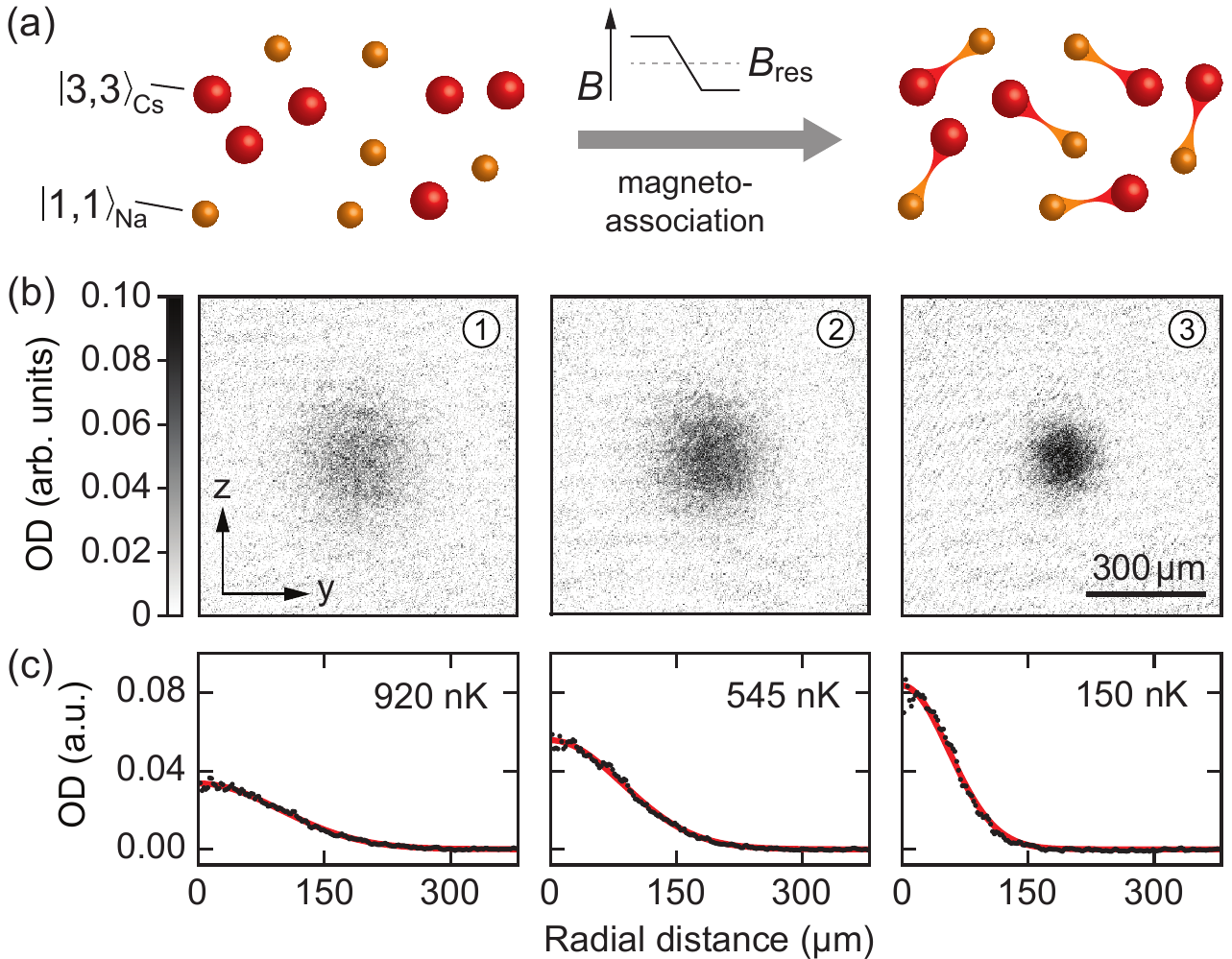}\\
    \caption{Formation of ultracold NaCs Feshbach molecules. (a) Illustration of association of free atoms into weakly bound Feshbach molecules. (b) Absorption images of NaCs molecules after 16~ms of time of flight expansion. Images corresponds to the data shown in Fig.~\ref{fig:N_and_T}. Phase-space density increases from left to right. Gravity points in the $-z$ direction. (c) Azimuthally averaged optical density profiles (black points) with Gaussian fits (red line). The molecule temperatures are obtained as discussed in the main text.}
    \label{fig:molecule}
\end{figure}

In this Letter, we demonstrate the creation of ultracold ensembles of NaCs Feshbach molecules from overlapping Bose-Bose mixtures of $^{23}$Na and $^{133}$Cs and observe phase-space densities near 0.1. We study molecule formation for different temperatures and atom numbers of the initial mixtures, associating molecules from thermal Na and Cs gases, from Na BECs overlapping with thermal Cs gases, and from overlapping BECs (see Fig.~\ref{fig:molecule}). This allows us to control the properties of molecular ensembles over a wide dynamic range, from up to 20,000 molecules at $2\,\mu$K to 600 molecules at $70\,$nK. We find a lifetime of about 6 ms for high-density NaCs ensembles ($\sim 10^{12}\,$cm$^{-3}$), providing excellent conditions for the transfer of ultracold NaCs Fesh\-bach molecules into the rovibrational ground state via STIRAP.


NaCs molecules are of special interest due to an exceptionally large dipole moment of $d= 4.6$ D~\cite{aymar2005calculation} in the rovibrational ground state. For NaCs, the effective range of dipole-dipole interactions, $a_\mathrm{d} = md^2/(8 \pi \epsilon_0 \hbar^2)$, can reach tens of micrometers, an order of magnitude larger than for NaK and two orders of magnitude larger than for KRb ($m$ denotes the molecular mass). In earlier work, NaCs molecules have been created from laser-cooled mixtures of Na and Cs~\cite{shaffer1999photoassociative, Kleinert2007, haimberger2009formation, zabawa2011formation} and single NaCs molecules have been created in optical tweezer traps~\cite{liu2018building,zhang2020forming,cairncross2021assembly}. The present work constitutes a step towards ultracold bulk gases of NaCs molecules, enabled by our recent demonstration of miscible and overlapping quantum gas mixtures of Na and Cs~\cite{warner2021overlapping}. Due to its large dipole moment, NaCs is a  promising candidate for the realization of novel many-body phases in strongly interacting dipolar quantum matter \cite{lahaye2009physics, baranov2012condensed, micheli2006toolbox, buchler2007strongly, Capogrosso2010}, making the pursuit of a BEC of NaCs ground state molecules an exciting goal.


Our experiment begins with the preparation of ultracold mixtures of Na and Cs in their hyperfine ground states $\ket{F, m_F} = \ket{1, 1}$ and $\ket{3,3}$, respectively. Here, $F$ denotes the total angular momentum and $m_F$ its projection on the magnetic field axis. Following evaporative cooling in the magnetic trap, the mixture is transferred into a crossed optical dipole trap operating at $1064\,$nm and further evaporatively cooled in the presence of a strong magnetic field of $B_\mathrm{prep} = 894\,$G. In the optical dipole trap, the trap frequencies of Na and Cs are very similar, $\omega^\mathrm{Na} \approx 1.08\,\omega^\mathrm{Cs}$, such that the differential gravitational sag is minimal. At the end of evaporation the trap frequencies are $\{\omega_x, \omega_y, \omega_z\} = 2 \pi \times \{74(1),54(1),332(7)\}\,$Hz (measured for Na). At $B_\mathrm{prep}$ the intra- and interspecies scattering lengths are such that Na and Cs clouds are miscible~\cite{warner2021overlapping}. 

\begin{figure} [t]
    \centering
    \includegraphics[width = 8.6 cm]{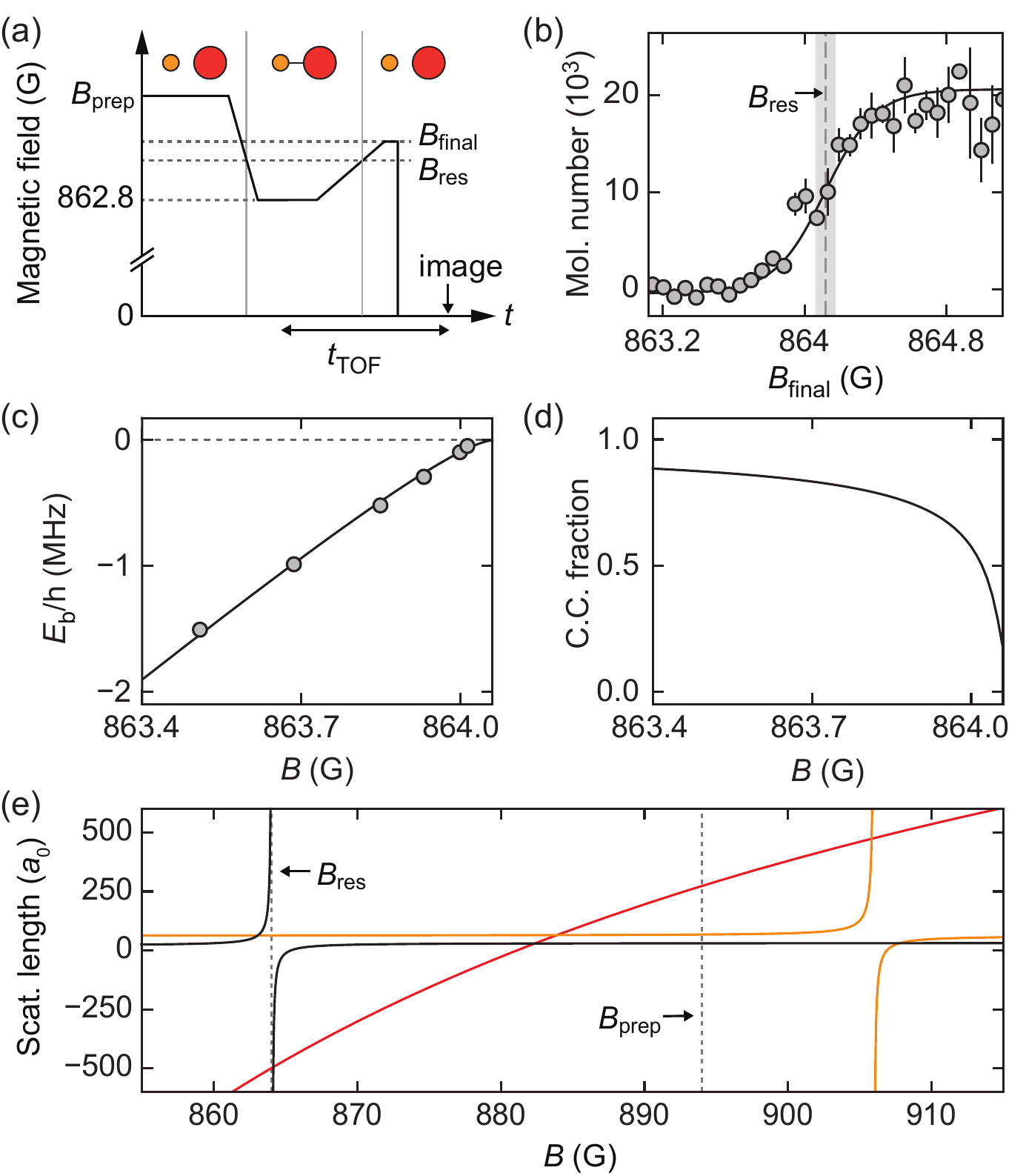}\\
    \caption{Magneto-association of NaCs Feshbach molecules and characterization of the Feshbach resonance. (a) Schematic of the magnetic field ramp for molecule association, dissociation, and subsequent imaging. $t_\mathrm{TOF}$ denotes the duration of time of flight expansion. (b) Molecule dissociation yields the accurate location of the Feshbach resonance. The gray-shaded region indicates $B_\mathrm{res}=864.12(5)\,$G. (c) Molecular binding energy as a function of magnetic field. The solid line shows a fit using a coupled-channel calculation. (d) Closed-channel fraction as a function of magnetic field, determined from the coupled-channel calculation. (e) Scattering length of Na-Na (orange)~\cite{knoop2011feshbach}, Cs-Cs (red)~\cite{berninger2013feshbach}, and Na-Cs (black) as a function of magnetic field. }
    \label{fig:rampandBE}
\end{figure}

First, we identify a Feshbach resonance for molecule formation. Using trap loss spectroscopy, we locate a Feshbach resonance at $864\,$G with an approximate width of $1\,$G~\cite{SI}. This resonance has been theoretically predicted~\cite{docenko2006coupling} and was recently observed in Ref.~\cite{zhang2020forming}. With the coarse location of the resonance known, we associate NaCs molecules via a magnetic field ramp from $B_\mathrm{prep}$ to $862.8\,$G [see Fig.~\ref{fig:rampandBE}(a)]. The ramp quickly traverses the region of increasingly attractive Cs-Cs scattering length [see Fig.~\ref{fig:rampandBE}(e)] with a speed of 15$\,$G/ms to suppress three-body losses of Cs, before crossing the Fesh\-bach resonance at a reduced speed of 1$\,$G/ms to ensure adiabatic molecule formation. After association, we apply a magnetic field gradient of 25$\,$G/cm and reduce the depth of the optical dipole trap by a factor of four. The gradient removes non-associated Na and Cs atoms from the trap, while levitating NaCs Feshbach molecules~\footnote{The magnetic moment of Na (Cs) is 0.5 (0.75)$\,\mu_\mathrm{B}$ and the magnetic moment of NaCs Feshbach molecules is -1.4(1)$\,\mu_\mathrm{B}$.}. This allows for background-free imaging of molecules (see details in the Supplemental Material~(SM)~\cite{SI}).

We determine the resonance position by measuring the magnetic field at which NaCs molecules dissociate using the magnetic field ramp shown in Fig.~\ref{fig:rampandBE}(a)~\cite{regal2004observation,zwierlein2004condensation}. If $B_\mathrm{final}$ is above the resonance, the molecules are  dissociated and detected; if $B_\mathrm{final}$ is below the resonance, they remain undetected. Figure~\ref{fig:rampandBE}(b) shows the determination of the resonance location at $B_\mathrm{res} = 864.12(5)\,$G. In addition, we measure the molecular binding energy as a function of magnetic field with the oscillating magnetic field method~\cite{thompson2005ultracold}. 
We record loss spectra at several magnetic field values (see details in SM~\cite{SI}) and obtain the data shown in Fig.~\ref{fig:rampandBE}(c). The binding energy shows a linear behavior up to very close to the resonance, which indicates that the molecular state is closed-channel dominated. This is confirmed by a coupled-channel calculation, which allows the calculation of the closed-channel fraction as shown in Fig.~\ref{fig:rampandBE}(d). The measured and calculated parameters are summarized in Table~\ref{tab:fbr} and show good agreement with Ref.~\cite{zhang2020forming}. The Na-Cs scattering length that results from the parametrization of the Feshbach resonance is displayed in Fig.~\ref{fig:rampandBE}(e).

\begin{table} [b]
    \centering
    \caption{Parametrization of the Feshbach resonance using $a(B) = a_\mathrm{bg} \left[ 1 - \Delta/(B - B_\mathrm{res}) \right]$. The error bar in the position reflects the systematic uncertainty in our magnetic field calibration. Theoretical values are obtained from a coupled-channel calculation.}
    \label{tab:fbr}
\begin{tabular}{c c c} \hline \hline
& Experiment & Theory  \\ \hline
$B_\mathrm{res}$ (G) & 864.12(5) & 864.07(2)  \\
$a_\mathrm{bg}$ ($a_0$) & 29(4) \,\cite{warner2021overlapping} & 31.58(3)  \\
$\Delta$ (G) & $\sim 1$& 1.30(1)   \\ \hline \hline
\end{tabular}
\end{table}


\begin{figure}
    \centering
    \includegraphics[width = 8.6 cm]{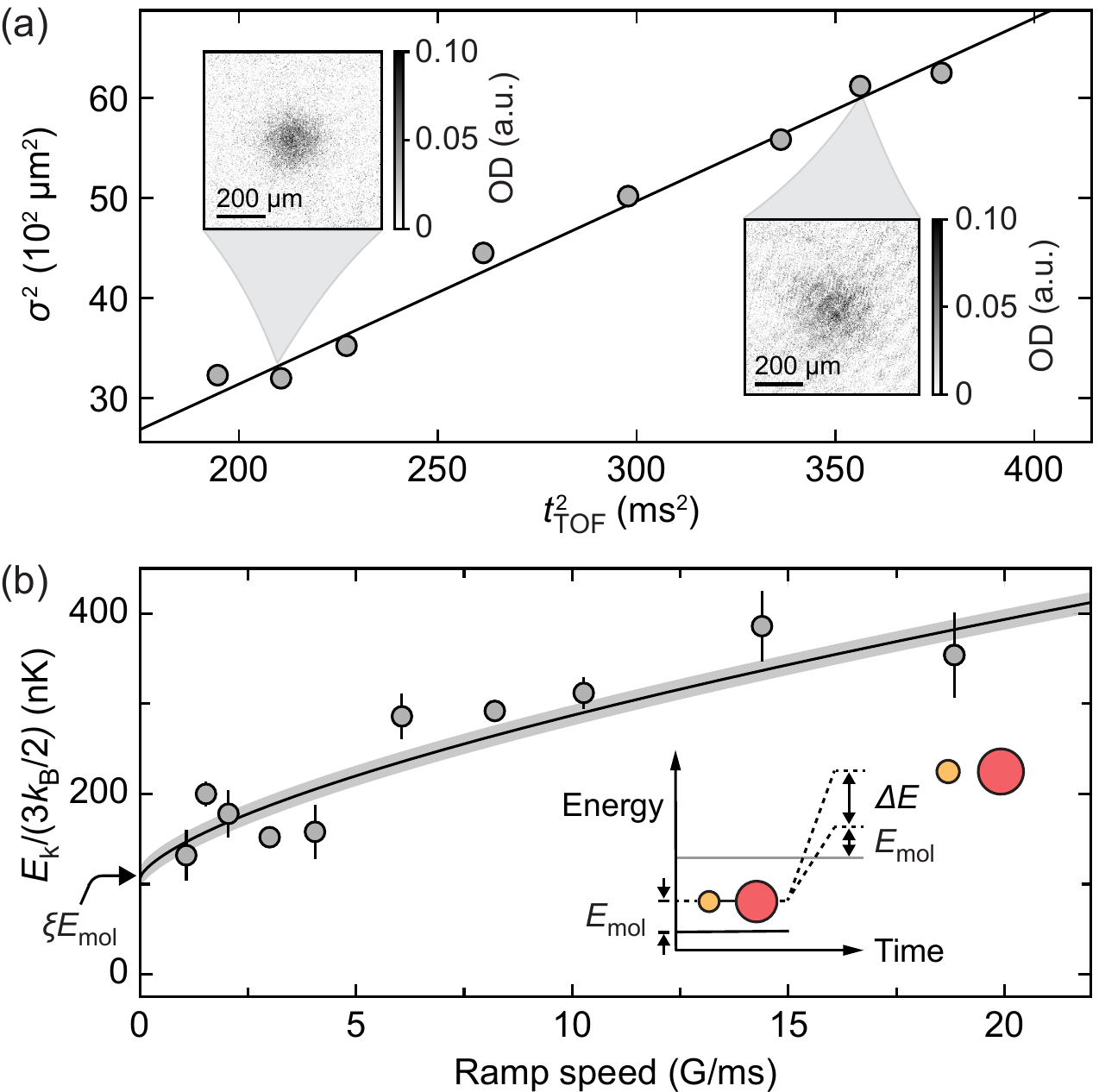}\\
    \caption{Measuring the mean kinetic energy of NaCs Feshbach molecules. (a) Cloud size $\sigma$ vs time of flight $t_\mathrm{TOF}$ for a dissociation ramp speed of $10\,$G/ms. $\sigma$ is obtained from a two-dimensional Gaussian fit to the Cs clouds after dissociation. The solid line shows a linear fit where the slope equals to $\Delta\sigma^2/\Delta \, t_\mathrm{TOF}^2$ (see text). (b) Mean kinetic energy of the resulting Cs cloud vs ramp speed. The solid line shows a fit of $\xi E_\mathrm{mol} + \Delta E (\dot{B})$, where $E_\mathrm{mol}$ is the only fitting parameter. For this data set, the fit yields $T_\mathrm{mol} = 124(13)$~nK.}
    \label{fig:time_of_flight}
\end{figure}

\begin{figure} [!t]
    \centering
    \includegraphics[width = 8.6 cm]{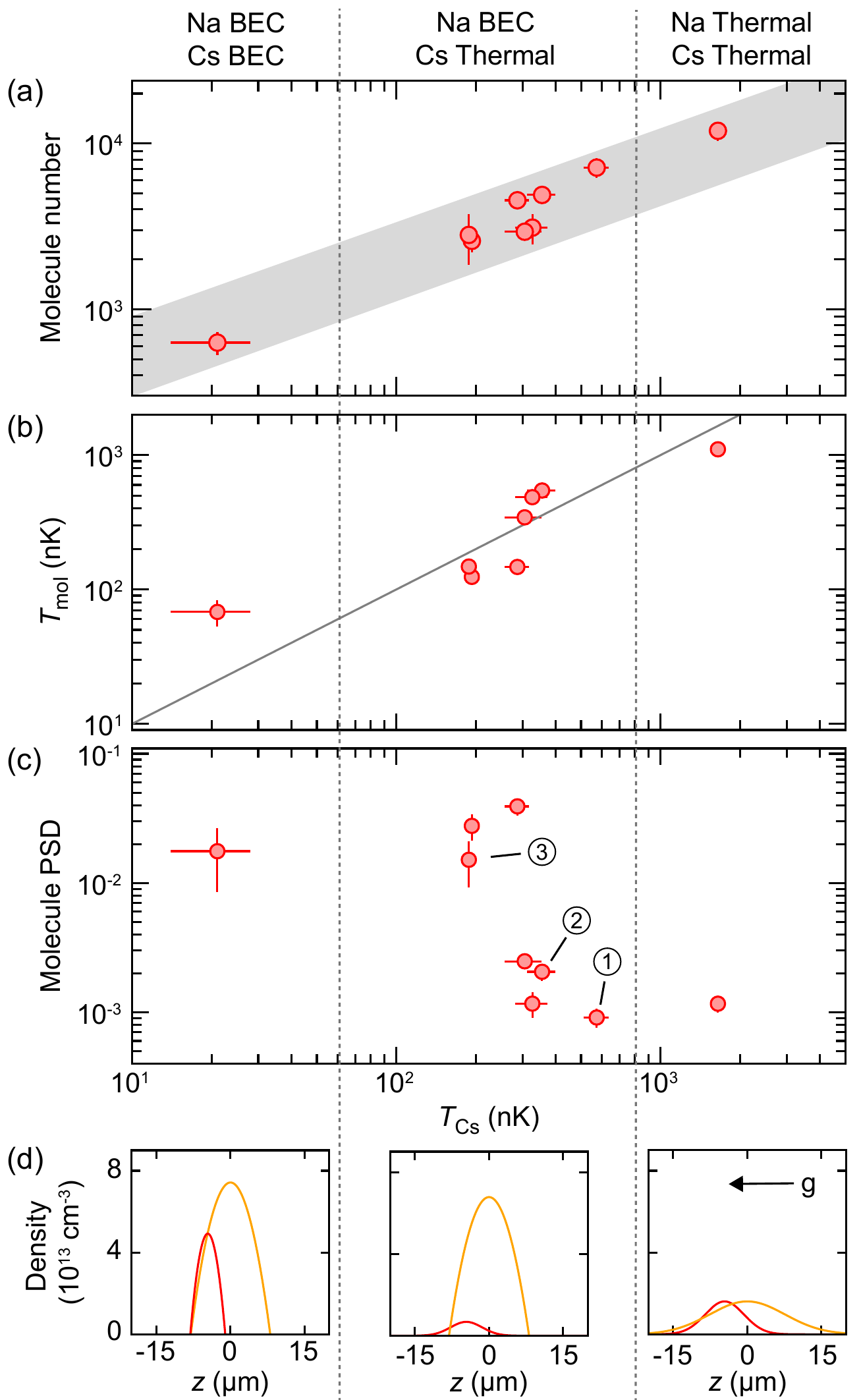}\\
    \caption{Creation of NaCs Feshbach molecules across a wide parameter range. Molecules are associated from overlapping thermal clouds of Na and Cs (right), overlapping Na BECs and Cs thermal clouds (middle), and overlapping Na and Cs BECs (left). (a) Molecule number vs Cs temperature, $T_\mathrm{Cs}$. The gray-shaded region indicates a fraction of $4(1)\,$\% of the initial Cs atom number. (b) Molecule temperature, $T_\mathrm{mol}$, vs $T_\mathrm{Cs}$. The solid line indicates $T_\mathrm{mol} = T_\mathrm{Cs}$. (c) Phase-space density of the molecular cloud vs $T_\mathrm{Cs}$. Numbered data points correspond to Fig.~\ref{fig:molecule}(b). (d) Density profiles and cloud overlap of Na (orange) and Cs (red). The shift of the cloud centers reflects the differential gravitational sag between Na and Cs.}
    \label{fig:N_and_T}
\end{figure}

We characterize the properties of the associated molecular clouds in terms of molecule number, temperature, and phase-space density. For the temperature measurement, we develop a protocol that is based on the time of flight expansion of dissociated molecules, following the sequence shown in Fig.~\ref{fig:rampandBE}(a). Right after association, the molecules are released from the trap, initiating time of flight expansion, and non-associated atoms are separated via a magnetic field gradient. After 5$\,$ms, the molecules are dissociated by a reverse field ramp with variable speed [see details in SM~\cite{SI}]. The dissociated molecules continue ballistic expansion as Na and Cs atoms for a variable time of flight until an image of Cs is taken [see Fig.~\ref{fig:time_of_flight}(a)]. From the expansion data we obtain the mean kinetic energy via  $E_\mathrm{k} =(3/2)m_\mathrm{Cs} \Delta \, \sigma^2/\Delta \, t_\mathrm{TOF}^2$~\cite{mukaiyama2004dissociation}, where $m_\mathrm{Cs}$ is the mass of a Cs atom.

To extract the mean kinetic energy of the molecules prior to dissociation, we take into account that $E_\mathrm{k}$ is comprised of the mean kinetic energy of the molecules $E_\mathrm{mol}$ and an additional kinetic energy from the dissociative reverse ramp, $\Delta E$. 
The additional kinetic energy arises from non-adiabaticity of the dissociation ramp and depends on the ramp speed and the coupling strength of the Feshbach resonance; the latter is characterized by the background scattering length, $a_\mathrm{bg}$, and the resonance width, $\Delta$. This has been previously considered for homonuclear molecules~\cite{mukaiyama2004dissociation, durr2004dissociation}. For our case, involving heteronculear molecules, we find that the additional kinetic energy carried by Cs depends on the ramp speed as~\cite{SI}
\begin{equation}
    \Delta E(\dot{B}) =  \frac{m_\mathrm{Na}}{m_\mathrm{Na} + m_\mathrm{Cs}} \Gamma\left(5/3\right) \left( \frac{3 \hbar^2 \dot{B} }{4 |\Delta| a_\mathrm{bg} \sqrt{2 \mu}} \right)^{2/3},
    \label{eq:ramp_energy}
\end{equation}
where $\Gamma$ is the Gamma function, $m_\mathrm{Na}$ is the mass of a Na atom, $\hbar$ is Planck's constant divided by $2\pi$, $\mu = m_\mathrm{Na} m_\mathrm{Cs}/(m_\mathrm{Na}+m_\mathrm{Cs})$ the reduced mass of the molecule, and $\dot{B}$ the ramp speed across the resonance. Using the fitted values for $a_{\rm bg}$ and $\Delta$ (see Table~\ref{tab:fbr}), we quantify $\Delta E$ without free parameters using Eq.~(\ref{eq:ramp_energy}) and find good agreement with our experimental data for different ramp speeds, as shown in Fig.~\ref{fig:time_of_flight}(b). From the fit we extract the energy $\xi E_\mathrm{mol}$ at zero ramp speed (full adiabaticity), where $ E_\mathrm{mol}$ is the mean kinetic energy of the molecular gas prior to dissociation and $ \xi = m_\mathrm{Cs}/(m_\mathrm{Na} + m_\mathrm{Cs})$.

In Fig.~\ref{fig:N_and_T}, we study the dependence of molecule number, temperature, and phase-space density~\cite{SI} on the conditions of the initial atomic mixtures. To this end, we associate molecules from thermal gases of Na and Cs, from Na BECs overlapping with a thermal gas of Cs, and from overlapping BECs of Na and Cs. For Na, the initial numbers range from $3.5\times 10^6$ to $5\times 10^5$, and, for Cs, from $9\times 10^5$ to $4\times 10^4$ atoms. The Na temperatures are typically slightly lower than the Cs temperatures due to a lag in thermalization of Cs during sympathetic cooling with Na~\cite{warner2021overlapping}. We use the Cs temperature $T_\mathrm{Cs}$ as a parameter that characterizes the starting conditions at $B_\mathrm{prep}$ prior to association.

We observe NaCs samples with $1.2(2)\times 10^4$ molecules when associated from thermal mixtures, and $6(1)\times 10^2$ molecules when associated from overlapping BECs [see Fig.~\ref{fig:N_and_T}(a)]. The detected molecule numbers correspond to $4(1)\,$\% of the Cs atoms prior to association, independent of the temperature of the initial mixture. While this fraction is similar to previous observations in heteronuclear bosonic  molecules~\cite{wang2015formation,koppinger2014production,voges2020formation}, it is likely not a fundamental limit. Using the pairing model of Ref.~\cite{hodby2005production} we expect that 80$\,$\% of the Cs atoms are associated into molecules, owing to well-matched densities and good cloud overlap [see Fig.~\ref{fig:N_and_T}(d)]. Indeed, we observe that the majority of Cs atoms vanishes during Feshbach association from dual BECs, suggesting that the molecule formation efficiency is high, but collisional loss between molecules and atoms during the relatively slow atom removal procedure ($\sim$ms) reduces the number of detectable molecules. A fast atom removal procedure via optical clean-out ($\sim\mu$s) is expected to improve detectable molecule numbers~\footnote{During completion of the manuscript, we have implemented a faster optical clean-out procedure and observe an enhancement of the detectable molecule number by at least a factor of 2.}.

The temperatures of the molecular clouds range from 2(1)$\,\mu$K to 68(15)$\,$nK [see Fig.~\ref{fig:N_and_T}(b)]. We calculate the temperature via $T_\mathrm{mol} = 3E_\mathrm{mol}/2 k_\mathrm{B}$, which serves as a measure of mean kinetic energy as we do not have direct evidence for thermalization of the molecular ensembles. The molecules follow the trend of the Cs temperatures, especially for the association from thermal mixtures and from Na BECs overlapping with a Cs thermal cloud. For the association from dual BECs, a marked increase of temperature compared to the initial Cs temperature is observed; we attribute this to heating of the Cs BEC during the ramp from $B_\mathrm{prep}$ to below $B_\mathrm{res}$, where the Cs scattering length varies from moderately repulsive to strongly attractive [see Fig.~\ref{fig:rampandBE}(e)]. Despite the heating, this is one of the lowest temperatures reported for heteronuclear bosonic Feshbach molecules.

Combining the data on molecule numbers and temperatures, we determine phase-space densities that range from $2.4(4) \times 10^{-4}$ up to $4(1)\times 10^{-2}$. We find the highest phase-space densities when molecules are associated from a Na BEC overlapping with a thermal Cs gas at the verge of condensation. The data demonstrate excellent starting conditions for the preparation of ultracold NaCs ground state molecules. Recently, effective evaporative cooling has been shown for ground state molecules using resonant collisional shielding~\cite{matsuda2020resonant,li2021tuning} and microwave shielding~\cite{anderegg2021observation,schindewolf2022evaporation}. For NaCs, resonant shielding is predicted to lead to a superb ratio of elastic to inelastic collisions of $10^6$~\cite{martinez2017adimensional,li2021tuning} and microwave shielding should be highly effective thanks to weak hyperfine interactions~\cite{karman2018microwave,lassabliere2018controlling}. This makes NaCs a promising candidate for direct evaporation to a BEC of dipolar ground state molecules, in particular when starting from a high initial phase-space density.

\begin{figure} [t]
    \centering
    \includegraphics[width = 8.6 cm]{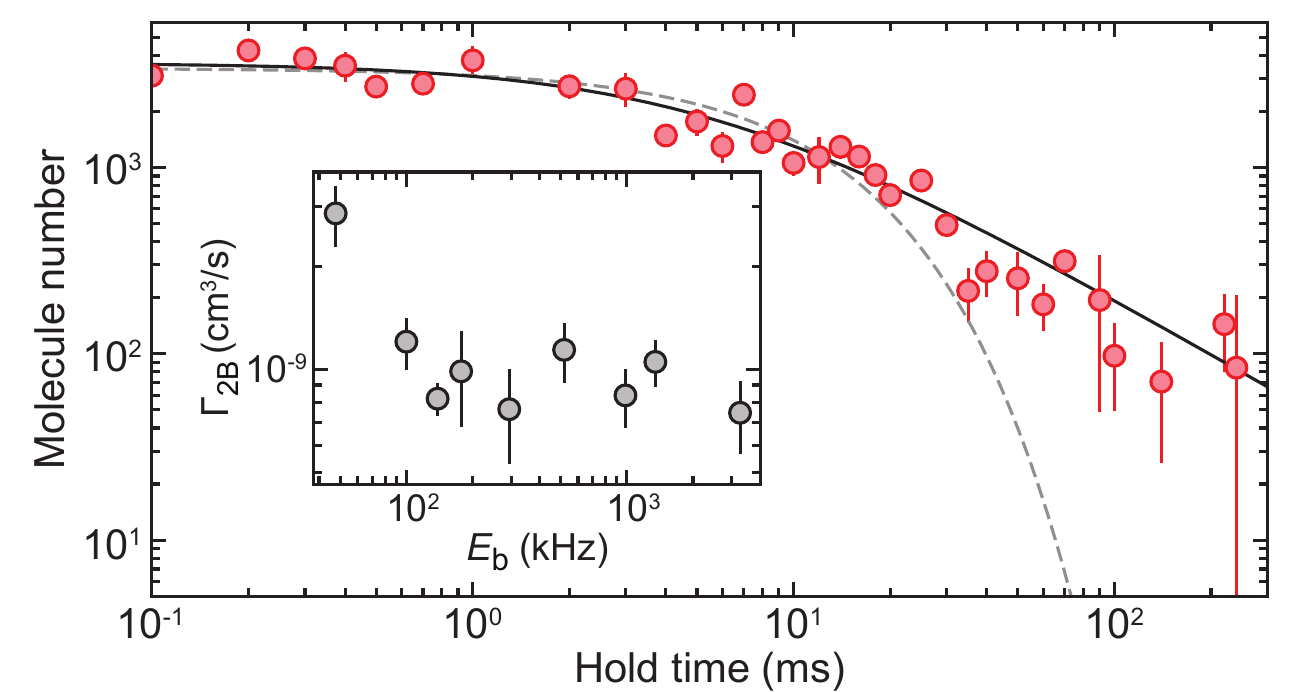}\\
    \caption{Lifetime of NaCs Feshbach molecules. Decay dynamics are recorded for molecules at a binding energy of $h \times 1.3\,$MHz. The solid (dashed) line is a fit to two-body (one-body) decay. The inset shows the derived two-body loss coefficients as a function of molecular binding energy for a cloud temperature of 1.8(4)$\, \mu$K. We attribute the increased decay at $E_\mathrm{b} = h \times 50\,$kHz to thermal dissociation of Feshbach molecules as the binding energy is comparable to the temperature.}
    \label{fig:Lifetime}
\end{figure}

Finally, we study the lifetime of NaCs molecular ensembles, while held in the vicinity of the Feshbach resonance. Figure~\ref{fig:Lifetime} shows a decay curve for a gas in the regime of highest phase-space density with a peak density of $n_0 = 8(2)\times10^{11}$ cm$^{-3}$ and a temperature of 147(9)$\,$nK. The data are consistent with two-body loss with a decay rate of $\Gamma_{\rm 2B}= 5(2)\times10^{-10}\,$cm$^{3}$s$^{-1}$ and a characteristic decay time of $1/(\Gamma_\mathrm{2B} \bar{n}) = 6(1)\,$ms, where $\bar{n}$ is the average density \footnote{The fit function for two-body decay has the form $N(t) = N_0/(1 + \Gamma_\mathrm{2B} \bar{n} t)$, where $N_0$ is the initial molecule number. The average density $\bar{n}$ of a thermal gas in a harmonic trap is given by $\bar{n} = n_0/2^{3/2}$, where $n_0$ is the peak density.}. We attribute the two-body decay to vibrational relaxation when two molecules collide. We do not observe a strong dependence of the decay rate on molecular binding energy (see the inset of Fig.~\ref{fig:Lifetime}). This is expected as the Franck-Condon factor for vibrational relaxation is proportional to the closed-channel fraction, which, as shown above, saturates for small binding energies (see Fig.~\ref{fig:rampandBE}). The observed lifetime is sufficiently long for fast STIRAP transfer ($\sim 100\,\mu$s) of a high phase-space density gas of NaCs molecules into the rovibrational ground state. 

In conclusion, we have demonstrated the creation of high phase-space density gases of NaCs Feshbach molecules, which is an ideal stepping-stone for the preparation of ultracold gases of NaCs ground state molecules. In the ground state, evaporative cooling supported by microwave~\cite{karman2018microwave,lassabliere2018controlling,anderegg2021observation,schindewolf2022evaporation} or resonant shielding~\cite{matsuda2020resonant,li2021tuning} may open the door to a BEC of dipolar NaCs molecules. In addition, mixtures of Na and Cs have favorable properties for the creation of dual Mott insulators in an optical lattice. This offers a path towards the efficient formation of Feshbach molecules~\cite{moses2015creation,reichsollner2017quantum} and low-entropy lattice gases of strongly dipolar NaCs ground state molecules.

We thank Rudolf Grimm and Tarik Yefsah for fruitful discussions and for helpful comments on the manuscript. We also thank Joseph Lee and Edita Bytyqi for experimental assistance. This work was supported by an NSF CAREER Award (Award No. 1848466), an ONR DURIP Award (Award No. N00014-21-1-2721) and a Lenfest Junior Faculty Development Grant from Columbia University. S.W. acknowledges additional support from the Alfred P. Sloan Foundation. I.S. was supported by the Ernest Kempton Adams Fund. C.W. acknowledges support from the Natural Sciences and Engineering Research Council of Canada (NSERC) and the Chien-Shiung Wu Family Foundation. W. Y. acknowledges support from the Croucher Foundation.

\end{document}


\title{Supplemental Material for ``A High-Phase Space Density Gas of NaCs Feshbach Molecules''}

\preprint{APS/123-QED}

\author{Aden Z. Lam}
\affiliation{Department of Physics, Columbia University, New York, New York 10027, USA}
\author{Niccol\`{o} Bigagli}
\affiliation{Department of Physics, Columbia University, New York, New York 10027, USA}
\author{Claire Warner}
\affiliation{Department of Physics, Columbia University, New York, New York 10027, USA}
\author{Weijun Yuan}
\affiliation{Department of Physics, Columbia University, New York, New York 10027, USA}
\author{Siwei Zhang}
\affiliation{Department of Physics, Columbia University, New York, New York 10027, USA}
\author{Eberhard Tiemann}
\affiliation{Institut f\"{u}r Quantenoptik, Leibniz Universit\"{a}t Hannover, 30167 Hannover, Germany}
\author{Ian Stevenson}
\affiliation{Department of Physics, Columbia University, New York, New York 10027, USA}
\author{Sebastian Will}
\affiliation{Department of Physics, Columbia University, New York, New York 10027, USA}

\date{\today}

\maketitle

\section{I: Coarse search of Feshbach resonance via trap loss spectra}
\begin{figure} [tbh]
    \includegraphics[width = 12 cm]{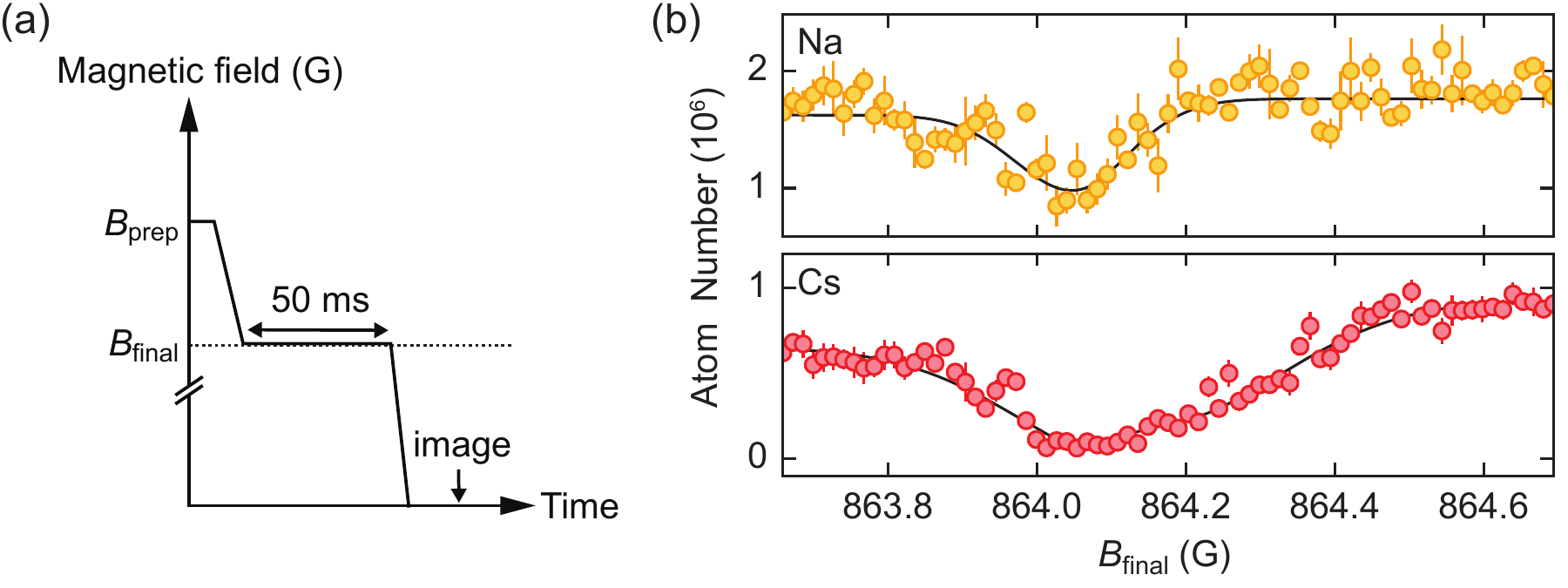}\\
    \caption{Locating the Feshbach resonance. (a) Magnetic field sequence. (b) Atom loss profiles for Na (top) and Cs (bottom). The solid line in the Na data shows a Gaussian fit. The solid line in the Cs data shows a fit that is the product of a Gaussian and a hyperbolic tangent. The lines serve as a guide to the eye. The temperature of Na (Cs) is approximately 0.7 (2)~$\mu$K prior to the ramp.}
    \label{fig:abg}
\end{figure}

We detect the coarse location of the Feshbach resonance between Na $|1,1\rangle$ and Cs $|3,3\rangle$ via a measurement of enhanced three-body loss, as shown in Fig.~\ref{fig:abg}. First, an ultracold mixture of Na and Cs is prepared at $B_\mathrm{prep}=894\,$G as described in Ref.~\cite{warner2021overlapping}. Then, the magnetic field is lowered within 10 ms to $B_\mathrm{final}$ where the atoms are held for 50$\,$ms, as shown in Fig.~\ref{fig:abg}(a). The atoms are released and undergo 20 ms of time-of-flight expansion before images of Na and Cs are taken at a low bias field of 2$\,$G, as described in Sec.~\ref{sec:lowfieldimaging}. From the loss profiles of both species (see Fig.~\ref{fig:abg}(b)), we find the approximate resonance position at 864.1(1)$\,$G. The magnetic field strength is calibrated by driving radio-frequency (rf) and microwave transitions between hyperfine states of Na in the relevant field range. The precision of the magnetic field calibration is $\pm 15\,$mG. 

\section{II: Imaging of Feshbach molecules}
\setlabel{II}{sec:lowfieldimaging}
For imaging, the Feshbach molecules are dissociated into Na and Cs atoms via a reverse magnetic field ramp. After dissociation, the trapping potentials are switched off and a magnetic field gradient is applied for $5\,$ms to spatially separate the molecules from the non-associated atoms while still at high field above the Feshbach resonance. Then the magnetic field is rapidly lowered to $2\,$G within $2\,$ms, followed by $5\,$ms hold time for oscillations to damp out. Finally, an absorption image of Cs is taken with a minimum time-of-flight of $12\,$ms. Imaging Cs, rather than Na, is advantageous due to a smaller expansion rate during time-of-flight and reduced energy transfer from the dissociation ramp, as discussed in the main text. 

As can be seen in the absorption images in Fig.~1(b) and Fig.~3(a), there is no notable distortion of the atomic clouds, which indicates the influence of residual magnetic field curvature during time-of-flight expansion at high field is minimal. We confirmed this by fitting the clouds with two-dimensional Gaussian fits. In addition, we have performed a classical Monte-Carlo simulation of the time-of-flight expansion that takes into account the actual magnetic fields from our coils that are traversed during time-of-flight expansion. We found that the cloud distortion is minimal and that there is no significant impact on the temperature measurement.


\section{III: Binding energy measurement via RF association}

\begin{figure} [t]
    \centering
    \includegraphics[width = 10.7 cm]{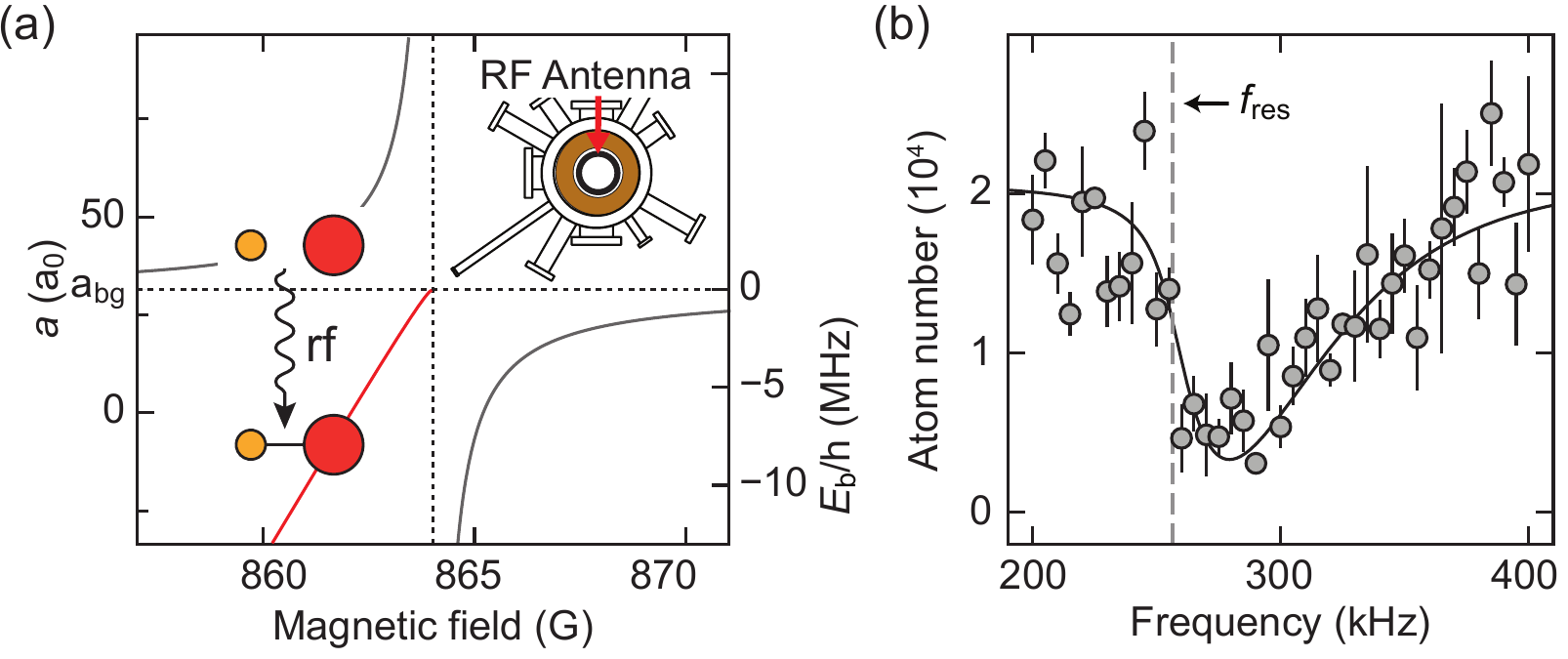}\\
    \caption{Oscillating magnetic field method to measure molecular binding energies. (a) When a radio frequency is applied in resonance with the molecular bound state, pairs of free atoms are transferred into the molecular state. The transfer appears as atom loss. (b) Measurement of the binding energy via an rf-pulse at 863.93~G, detected via atom loss in Cs. The solid black line shows a fit using Eq.~(\ref{eq:Bolt_Lorez}), where $\Gamma$, $f_\mathrm{res}$, $T_\mathrm{Cs}$ are free fit parameters. The dashed grey line indicates the position of $f_\mathrm{res}=E_\mathrm{b}/h$.}
    \label{fig:bindingenergy}
\end{figure}

We measure the binding energy $E_\mathrm{b}$ of the molecular bound state on the repulsive side of the Feshbach resonance via the oscillating magnetic field method (see Fig.~\ref{fig:bindingenergy}). An rf-pulse is applied to the mixture of Na and Cs atoms, as illustrated in Fig.~\ref{fig:bindingenergy}(a). When the oscillation frequency matches the molecular binding energy, transitions between free atoms and the molecular state are induced. Resonant transfer is detected as loss, as Feshbach molecules are quickly lost from the trap due to inelastic collisions. The rf field is produced by a single loop antenna placed concentrically with the magnetic field coils, powered by a 1$\,$Watt RF amplifier (Minicircuits ZHL-32A).

The rf frequency $f$ is varied while the magnetic field is kept constant, giving rise to loss spectra as shown in Fig.~\ref{fig:bindingenergy}(b) from which we extract the resonance frequency. The rf-pulse duration is typically 100$\,$ms. The loss spectra are inhomogeneously broadened due to the finite temperature of the atomic mixture of about $2\,\mu$K. We fit the Cs loss spectra with a convolution of a Boltzmann distribution and a Lorentzian~\cite{jones1999fitting},
\begin{equation}
    W(f, f_\mathrm{res}) \propto \int_0^\infty L_\Gamma (f, f_\mathrm{res} - E_\mathrm{kin}/h) e^{-\frac{E_\mathrm{kin}}{k_\mathrm{B} T_\mathrm{Cs}}} \sqrt{E_\mathrm{kin}} \, dE_\mathrm{kin},
    \label{eq:Bolt_Lorez}
\end{equation}
where $E_\mathrm{kin}$ is the kinetic energy, $k_\mathrm{B}$ is the Boltzmann constant, $L_\Gamma(f, f_0)$ is a the Lorentzian function with center frequency $f_0$ and linewidth $\Gamma$. We take the center of the Lorentzian as $E_\mathrm{b}=hf_\mathrm{res}$. 

\section{IV: Measurement of molecule temperature prior to dissociation
\label{app:rampspeed}}
When Feshbach molecules are dissociated via the reverse magnetic field ramp, the non-adiabaticity of the ramp causes the molecular state to rise above the energy of the free atom continuum. As a result, the molecular state is unstable and decays spontaneously into a pair of unbound atoms. The more non-adiabiatic the reverse ramp, the  higher the amount of added kinetic energy. The additional energy from non-adiabatic dissociation has previously been calculated in Refs.~\cite{mukaiyama2004dissociation} and \cite{durr2004dissociation} for homonuclear Na$_{2}$ and Rb$_{2}$ molecules, respectively. Here, we adapt the calculation to the case of heteronuclear NaCs molecules. The decay rate can be approximated by Fermi's Golden Rule,
 \begin{equation}
     P_\mathrm{if} = \frac{2 \Delta B \Delta \mu a_{bg} }{\hbar^2} \sqrt{2 \mu \Delta E},
 \end{equation}
where $\Delta B$ is the width of the Feshbach resonance, $\Delta \mu$ is the difference in magnetic moment between the molecule and the two free atoms, and $a_{\mathrm{bg}}$ is the background scattering length. Assuming a linear ramp through the resonance, the additional kinetic energy of Cs after dissociation takes the form
\begin{equation}
    \Delta E(\dot{B}) = \frac{m_\mathrm{Na}}{m_\mathrm{Na}+m_\mathrm{Cs}}\Gamma\left(5/3\right) \left( \frac{3 \hbar^2 \dot{B} }{4 |\Delta B| a_{bg} \sqrt{2 \mu}} \right)^{2/3},
\end{equation}
where $m_\mathrm{Na(Cs)}$ is the mass of a Na (Cs) atom, $\Gamma$ is the Gamma function, $\dot{B}$ is the ramp speed of the magnetic field, and $\Delta B$ is the width of the Feshbach resonance. The prefactor $m_\mathrm{Na}/(m_\mathrm{Na}+m_\mathrm{Cs})$ originates from momentum conservation. 

In the experiment we measure the mean kinetic energy $E_\mathrm{k}$ of Cs after dissociation. The mean kinetic energy of the molecule prior to dissociation is given by,
\begin{gather}
    E_\mathrm{mol} = \left(E_\mathrm{k} - \Delta E (0) \right )/\xi,
\end{gather}
where $\xi=m_\mathrm{Cs}/(m_\mathrm{Na}+m_\mathrm{Cs})$. The equivalent temperature is calculated via $T_\mathrm{mol}=2 E_\mathrm{mol}/ 3 k_\mathrm{B}$ from the equipartition theorem.

\section{V: Measurement of mean kinetic energy versus ramp speed}

\begin{figure} [tbh]
    \centering
    \includegraphics[width = 6 cm]{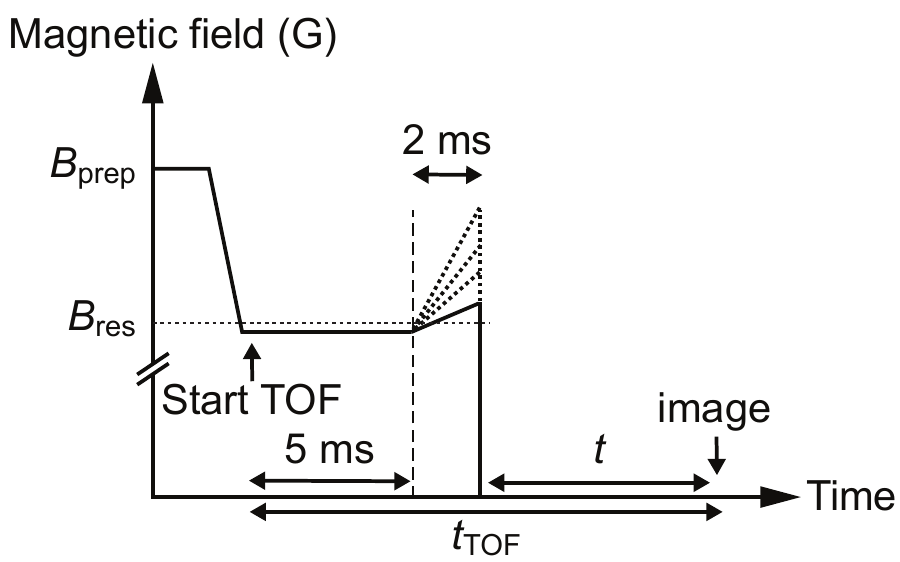}\\
    \caption{Magnetic field sequence for the measurement of $E_\mathrm{k}$.}
    \label{fig:rampspeed}
\end{figure}

The reverse field ramp used to characterize the mean kinetic energy of the molecules is illustrated in Fig.~\ref{fig:rampspeed}. The ramp speed, $\dot{B}$  is varied by fixing the ramp duration and changing the final magnetic field it ramps to.
Time-of-flight measurements of the dissociated molecules are obtained by varying $t$. Typical data is shown in Fig.~3(a) in the main text. All temperature measurements are performed at sufficiently long times-of-flight, such that the expansion is in the linear regime, $\sigma \approx v_\mathrm{rms} t_\mathrm{TOF}$. The linear slope yields the root-mean-square velocity $v_\mathrm{rms}$, from which the mean kinetic energy is calculated via $E_\mathrm{k} = 3m_\mathrm{Cs} v_\mathrm{rms}^2/2$.

\section{VI: Calculation of Molecular Phase-Space density}

From the molecule number, $N_\mathrm{mol}$, and temperature, $T_\mathrm{mol}$, we obtain the corresponding phase-space density in a harmonic trap via 
\begin{gather}
    \mathrm{PSD_{mol}} = N_\mathrm{mol} \left( \frac{2\pi\hbar}{k_\mathrm{B}T_\mathrm{mol}} \right) ^3 \omega^\mathrm{NaCs}_x \omega^\mathrm{NaCs}_y \omega^\mathrm{NaCs}_z.
\end{gather}
The trap frequencies for the Feshbach molecules are obtained from the trap frequencies of Na and Cs, assuming that the polarizability equals the sum of the atomic polarizabilities and the mass equals to the sum of the atomic masses~\cite{vexiau2017dynamic}. For our trapping conditions, the trapping frequencies of NaCs Feshbach molecules are related to the Na trapping frequencies as $\omega^\mathrm{NaCs}\approx 1.04 \omega^\mathrm{Na}$. We have cross-checked this by directly measuring the trap frequencies with the Feshbach molecules.